# Long-Term Stability of Superconducting Metal Superhydrides

**Comment on "Diffusion-Driven Transient Hydrogenation in Metal Superhydrides at Extreme Conditions"** [Y. Zhou, Y. Fu, M. Yang et al. in *Nat. Commun.* **16**, 1135 (2025)]

Vasily S. Minkov[1*], Mikhail A. Kuzovnikov[2], Panpan Kong[3,4], Alexander P. Drozdov[1], Feng Du[1], Jiafeng Yan[1,5], Jaeyong Kim[5], Stella Chariton[6], Vitali B. Prakapenka[6], Mohamed Mezouar[7], Björn Wehinger[7], G. Alexander Smith[8], Fedor F. Balakirev[8], Evgeny F. Talantsev[9,10]

[1]*Max-Planck Institute for Chemistry, Mainz, Germany*
[2]*Centre for Science at Extreme Conditions and School of Physics and Astronomy, University of Edinburgh, Edinburgh, U.K.*
[3] *Beijing National Laboratory for Condensed Matter Physics, Institute of Physics, Chinese Academy of Sciences, Beijing, China*
[4]*School of Physical Sciences, University of Chinese Academy of Sciences, Beijing, China*
[5]*Institute for High Pressure Research, Department of Physics, Hanyang University, Seoul, Republic of Korea*
[6]*Center for Advanced Radiation Sources, University of Chicago, Chicago, Illinois, USA*
[7]*European Synchrotron Radiation Facility, Grenoble, France*
[8]*NHMFL, Los Alamos National Laboratory, Los Alamos, New Mexico, USA*
[9]*Mikheev Institute of Metal Physics, Ekaterinburg, Russian Federation*
[10]*Ural Federal University, Ekaterinburg, Russian Federation*
*E-mail: v.minkov@mpic.de*

*Zhou et al., in their recent publication (Nat. Commun.* **16**, *1135, 2025), reported the synthesis of lanthanum superhydride, LaH$_x$ (x = 10.2–11.1), by laser heating LaH$_3$ with NH$_3$BH$_3$ at a pressure of 170 GPa and investigated the temporal evolution of the NMR spectra of the reaction products. They observed a gradual decrease in the $^1$H-NMR signal intensity assigned to the synthesized metal hydride, accompanied by an increase in molecular hydrogen within the sample chamber over a period of 50 days. Based on these observations, the authors concluded that LaH$_{10}$ progressively decomposes into LaH$_3$ and H$_2$ within two months after synthesis at its formation pressure of 170 GPa. Here, we demonstrate that, under their formation conditions, metal superhydrides are thermodynamically more stable than metal trihydrides. Furthermore, we present direct experimental evidence – based on X-ray diffraction and four-probe electrical resistance measurements – confirming the stability of both the crystal lattice and high-temperature superconducting properties of the Fm-3m-LaH$_{10}$ phase for more than five years. This long-term stability is consistent with predictions from quantum chemistry calculations.*

Numerous theoretical studies predict that, under sufficiently high pressure, metal superhydrides MH$_x$ (M = La, Y, Ce, Ca, etc., x ≥ 6) become stable with respect to decomposition into hydrogen and lower hydrides. For example, theoretical predictions for the binary yttrium-hydrogen and lanthanum-hydrogen systems indicate that *Im-3m*-YH$_6$ is thermodynamically stable above 110 GPa[1-3], while *Fm-3m*-LaH$_{10}$ is stable above 150 GPa[3]. Subsequent experiments confirmed the formation of *Im-3m*-YH$_6$ at 160-180 GPa[4-6] and *Fm-3m*-LaH$_{10}$ at 140-180 GPa[7-13], in reasonable agreement with the theoretical predictions.

Some metal superhydrides readily form in a hydrogen atmosphere at sufficiently high pressures at room temperature over the course of several weeks. However, laser heating accelerates the hydrogenation reaction at higher temperatures. When alternative hydrogen sources, such as NH$_3$BH$_3$, are used instead of H$_2$, laser heating becomes essential to decompose ammonia borane and facilitate the release of free H$_2$.

The formation of superconducting *Fm-3m*-LaH$_{10}$, *P6$_3$/mmc*-YH$_9$, *Im-3m*-YH$_6$, and *Im-3m*-YD$_6$ phases by exposing corresponding metal trihydrides/trideuterides to hydrogen/deuterium at high pressures and room temperature has been previously reported[5,10]. These metal superhydrides were characterized by X-ray powder diffraction and four-probe electrical resistance measurements, demonstrating high $T_c$ values, as shown in Figure 1.

Our experimental results contradict the findings of Zhou *et al.*[14], which suggest that lanthanum superhydride is unstable and decomposes into LaH$_3$ and H$_2$ at the pressures at which it was formed. In contrast, our experiments show that the chemical reaction proceeds in the opposite direction – *forming* superhydrides rather than

*decomposing* them. This provides direct experimental evidence that lanthanum and yttrium superhydrides are thermodynamically favored at high pressures. Consequently, these phases cannot spontaneously decompose into LaH₃ while releasing free H₂ at the conditions under which they were synthesized.

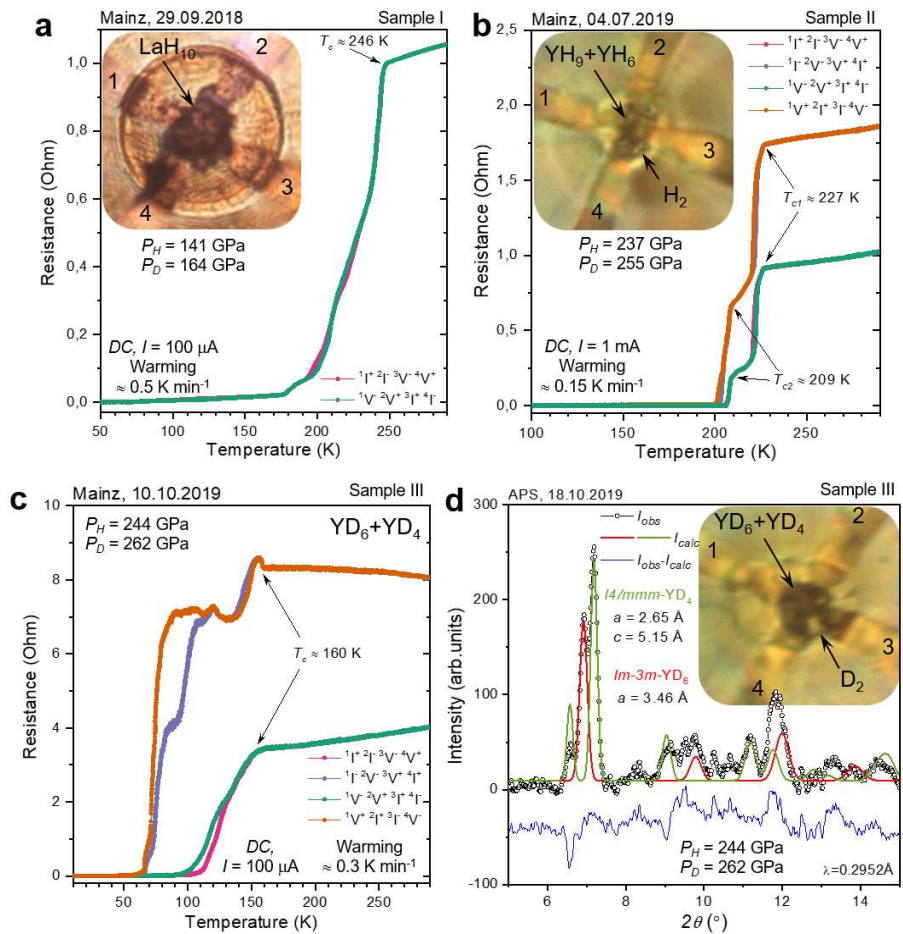

**Figure 1. Formation of metal superhydrides with high-temperature superconductivity at high pressures and room temperature. a–c)** Temperature dependence of the electrical resistance *R(T)* of metal superhydrides, demonstrating high-temperature superconductivity. **a)** Sample I: LaH₃ and H₂ were pressurized to $P_H$ = 141 GPa (estimated using H₂ (D₂) Raman vibron scale[15]) or $P_D$ = 164 GPa (estimated using diamond edge Raman scale[16]) on 12.09.2018. After 17 days at room temperature, *R(T)* measurements showed $T_c \approx$ 246 K (*Fm-3m*-LaH₁₀). **b)** Sample II: YH₃ and H₂ were pressurized to $P_H$ = 237 GPa ($P_D$ = 255 GPa) on 19.06.2019. After 15 days at room temperature, *R(T)* measurements showed $T_{c1} \approx$ 227 K (*P6₃/mmc*-YH₉) with a second step at $T_{c2} \approx$ 209 K (*Im-3m*-YH₆). X-ray powder diffraction pattern and the Rietveld refinement of Sample II are shown in Supplementary Figure 2 in Ref.[5] **c)** Sample III: YD₃ and D₂ were pressurized to $P_H$ = 244 GPa ($P_D$ = 262 GPa) on 23.09.2019. After 17 days at room temperature, *R(T)* measurements showed $T_c \approx$ 160 K (*Im-3m*-YD₆). **d)** X-ray powder diffraction pattern of Sample III and the Rietveld refinement confirms the presence of *Im-3m*-YD₆ and *I4/mmm*-YD₄ phases. The black circles are experimental data, the red and green profiles show the simulated contribution of the *Im-3m*-YD₆ and *I4/mmm*-YD₄ phases, and the blue profile represents the refinement residual.

Next, we address the long-term stability of metal superhydrides, a key focus of our research since the discovery of near-room-temperature superconductivity in highly compressed hydrides. To this end, we have systematically monitored several synthesized superhydride phases over years.

It is important to emphasize that the typical time interval between sample synthesis and subsequent resistance *R(T)* measurements, including those performed under external magnetic fields, ranges from several days to weeks[5,9,10], and can extend to several months for magnetic susceptibility measurements[8,17]. Furthermore, synchrotron X-ray diffraction experiments are often conducted weeks or even months after sample synthesis.

The authors of Ref.[14] acknowledge that the superconducting $Im\text{-}3m$-$H_3S$ phase remains stable and retains its $T_c$ of approximately 200 K over several years[18-21]. Below, we present evidence demonstrating that the $Fm\text{-}3m$-$LaH_{10}$ phase preserves its crystal lattice and high-temperature superconducting properties for over five years.

Figure 2 illustrates the stability of the $Fm\text{-}3m$-$LaH_{10}$ crystal lattice over at least 55 months in sample IV, as confirmed by X-ray diffraction experiments. Sample IV was synthesized from a La piece and $H_2$ using pulsed laser heating at $P_H$ = 156 GPa ($P_D$ = 172 GPa) on February 17, 2020. Nine days later, an initial X-ray diffraction experiment at the Advanced Photon Source (Argonne) revealed that the sample primarily consisted of the $Fm\text{-}3m$-$LaH_{10}$ phase, with a minor impurity of the $P6_3/mmc$-$LaH_{\sim10}$ phase, commonly observed in samples containing $Fm\text{-}3m$-$LaH_{10}$[9,10]. Four-probe electrical resistance measurements were conducted on December 27, 2022 – 34 months after synthesis – and revealed a superconducting transition at $T_c \approx$ 247 K. Fifty-five months after synthesis, a second X-ray diffraction experiment at the European Synchrotron Radiation Facility (Grenoble) confirmed that the sample retained the same phase composition and, therefore, the same chemical composition. A minor decrease (<0.5%) in the lattice parameters for both cubic and hexagonal phases was attributed to a slight pressure increase in the diamond anvil cell over time (see Figure 2).

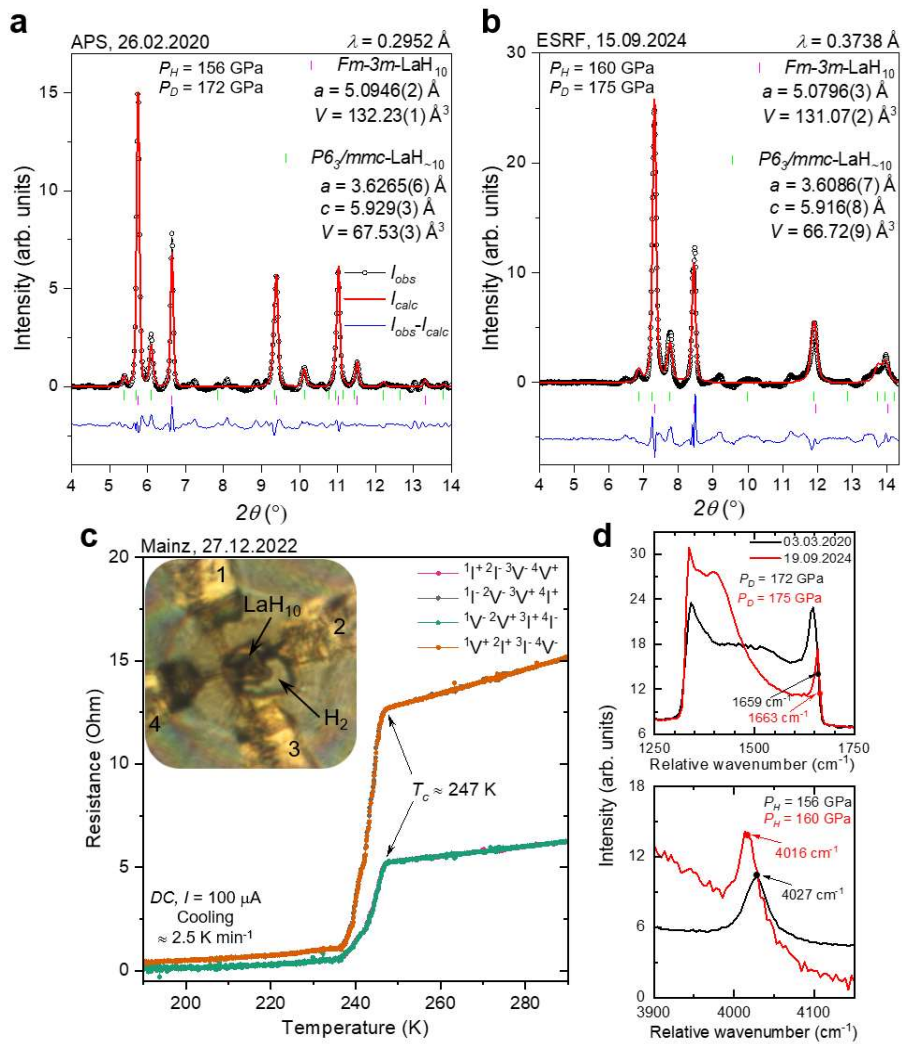

**Figure 2. Long-term stability of the $Fm\text{-}3m$-$LaH_{10}$ crystal lattice. a)** and **b)** X-ray powder diffraction patterns for sample IV were recorded one week and 55 months after the synthesis, respectively. The black circles represent the experimental data, while the red and blue profiles correspond to simulated patterns and the Le Bail refinement residuals, respectively. The magenta and green ticks indicate the calculated positions of the diffraction peaks for the $Fm\text{-}3m$-$LaH_{10}$ and $P6_3/mmc$-$LaH_{\sim10}$ phases. **c)** $R(T)$ measurements performed 34 months after synthesis, showing a superconducting transition with a $T_c \approx$ 247 K. **d)** Raman spectra, demonstrating the evolution of the estimated pressure in the diamond anvil cell over time, as manifested in the shift of the diamond Raman line edge (top) and the hydrogen vibron (bottom).

Figure 3 summarizes the experimental data for sample V, synthesized from a La piece and $H_2$ using laser heating at $P_H$ = 151 GPa ($P_D$ = 182 GPa) on September 21, 2018. Subsequent $R(T)$ measurements revealed a superconducting transition at $T_c \approx$ 248 K (see Figure 3a), and X-ray diffraction confirmed that the sample contained a dominant $Fm$-$3m$-$LaH_{10}$ phase with impurities of the $P6_3/mmc$-$LaH_{\sim 10}$ and $Pm$-$3m$-$LaH_{11-12}$ phases (see Figure 3d).

A crucial experimental observation is the resistive $R(T)$ data in Figure 3, showing that the superconducting transition was consistently reproduced 8 and 66 months after synthesis, with $T_c$ values remaining around 250 K. Minor deviations in $T_c$ are likely because of slight pressure variations in the diamond anvil cell over time.

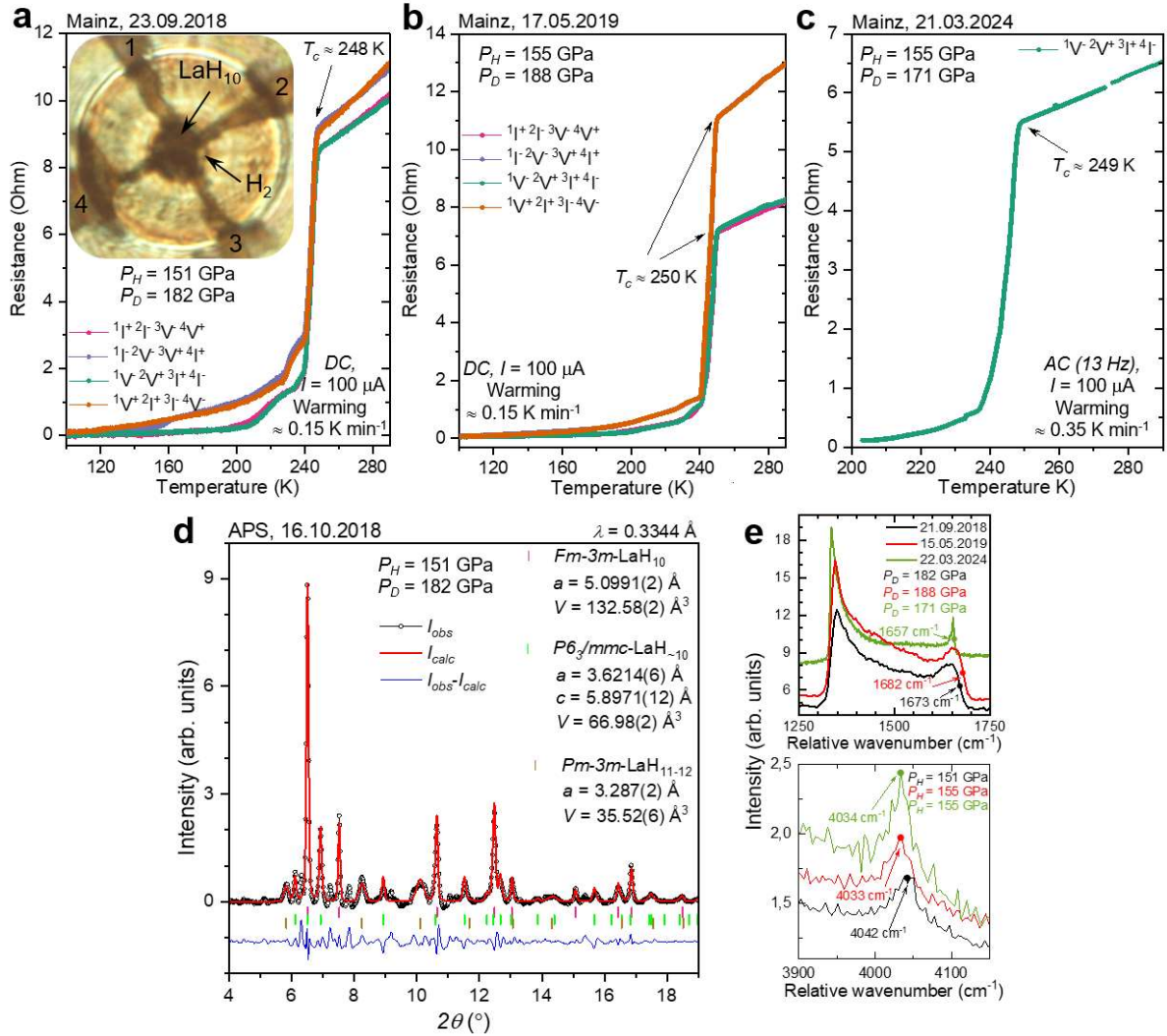

**Figure 3. Long-term stability of the superconducting properties of the $Fm$-$3m$-$LaH_{10}$ phase. a – c)** Electrical resistance $R(T)$ measurements for sample V taken respectively 2 days, 8 months, and 66 months after synthesis, showing a superconducting transition at approximately 250 K. **d)** X-ray powder diffraction patterns for sample V collected about one month after the synthesis. The black circles represent the experimental data, while the red and blue profiles correspond to the simulated pattern and the Le Bail refinement residuals, respectively. The magenta, green, and brown ticks indicate the calculated positions of the diffraction peaks for the $Fm$-$3m$-$LaH_{10}$, $P6_3/mmc$-$LaH_{\sim 10}$, and $Pm$-$3m$-$LaH_{11-12}$ phases, respectively. **e)** Raman spectra, demonstrating the evolution of the estimated pressure in the diamond anvil cell over time, as manifested in the shift of the diamond Raman line edge (top) and the hydrogen vibron (bottom).

Another significant experimental fact is that the $LaH_{10}$ phase remained intact even in the pressure range of 120-150 GPa[9,13], where the $Fm$-$3m$-$LaH_{10}$ phase undergoes minor structural distortion into the $C2/m$-$LaH_{10}$ phase[9]. This sample synthesized at $P_H$ of 138 GPa via laser heating on October 30, 2018, exhibited $T_c \approx$ 243 K in $R(T)$ measurements conducted on the following day. X-ray diffraction experiments at APS confirmed the presence of the dominant $Fm$-$3m$-$LaH_{10}$ phase on February 01, 2019. During transportation to the Los Alamos Magnet Laboratory, the sample pressure dropped to 120 GPa. Subsequent $R(T,H)$ measurements on April 19, 2019, and X-ray diffraction on May 31, 2019, revealed monoclinic distortions in the $Fm$-$3m$-$LaH_{10}$ phase and a significantly lower $T_c \approx 191$ K. However, recompression to 136 GPa on June 19, 2019, restored $T_c \approx 241$ K, and the sample´s superconducting transition was studied under high magnetic fields at Los Alamos until July 26, 2019. This study was published in Ref.[9]

These results demonstrate that even decompression down to 120 GPa does not lead to the decomposition of $LaH_{10}$ into phases with lower hydrogen content over several months. Thus, the contradictions with Ref.[14] cannot be attributed solely to differences in a pressure scale used in their experiments.

Furthermore, it is important to note that no corroborating measurements, such as X-ray diffraction, were performed to verify the presence of the $Fm$-$3m$-$LaH_{10}$ phase in the samples analyzed via NMR in Ref.[14] This is a critical issue, as NMR alone cannot unambiguously identify the composition of a sample.

In addition to the NMR study, the authors[14] reported electrical resistance measurements for four samples from a separate set of experiments. However, all samples exhibited very broad resistive transitions with $T_c$ values significantly below 250 K, indicating the absence of the $Fm$-$3m$-$LaH_{10}$ phase: two samples showed $T_c < 160$ K, while other two had $T_c < 80$ K. Likewise, these samples were not structurally characterized, leaving their phase compositions uncertain.

Ref.[14] also describes a sample in which $T_c$ dropped from approximately 150 K to 80 K within a few days, attributing this to hydrogen desorption from the superconducting phase. However, this argument is flawed, as an increase in hydrogen content can also result in a decrease in $T_c$ under certain conditions. For example, the formation of lanthanum dihydride is associated with a decrease of $T_c$ from approximately 6 K in pure lanthanum[22] to below 0.3 K in the metal dihydride[23].

In summary, we have presented experimental evidences demonstrating that high-temperature superconducting metal superhydrides, namely $Im$-$3m$-$YH_6$ and $Fm$-$3m$-$LaH_{10}$, are more thermodynamically stable under their formation conditions than their corresponding metal trihydrides, in agreement with theoretical predictions. Furthermore, the hydrogen-rich $Fm$-$3m$-$LaH_{10}$ phase has been found to remain stable for at least 66 months after synthesis, maintaining both its crystal lattice and superconducting properties ($T_c$), as confirmed by X-ray diffraction and four-probe electrical resistance measurements.

**Acknowledgements**

V.S.M., A.P.D., and F.D. express their gratitude to the Max Planck community for their support and to Prof. Dr. U. Pöschl for his continuous encouragement. The authors also thank Dr. Danila A. Barskiy for valuable discussions. M.A.K. acknowledges funding from the European Research Council under the European Union's Horizon 2020 Research and Innovation Program (Grant Agreement 948895, "MetElOne"). E.F.T. acknowledges support from the Ministry of Science and Higher Education of the Russian Federation within the framework of the state assignment for the IMP UB RAS, as well as research funding from the Ministry of Science and Higher Education of the Russian Federation under Ural Federal University Development Program (Priority-2030).

The authors acknowledge the Advanced Photon Source (APS) at Argonne National Laboratory and the European Synchrotron Radiation Facility (ESRF) for providing synchrotron radiation facilities at beamline GeoSoilEnviro CARS, Sector 13 (proposals GUP-58419, GUP-65644, GUP-67864) and beamline ID-27 (proposal HC-5940). The National High Magnetic Field Laboratory is supported by the National Science Foundation through NSF/DMR-2128556, the State of Florida, and the U.S. Department of Energy.

The authors also recognize the contribution of Dr. Mikhail I. Eremets, including the sample preparation, electrical resistance and X-ray diffraction measurements, and discussions of the results.

**Author contribution**

V.S.M., M.A.K., P.K., A.P.D., and F.D. prepared the samples and conducted electrical resistance measurements. F.F.B. performed electrical resistance measurements under high magnetic fields. V.S.M., F.D., J.Y., G.A.S., and J.K.

collected X-ray diffraction data with assistance from V.B.P., S.C., M.M., and B.W. V.S.M. and M.A.K. processed the structural data. V.S.M., M.A.K., E.F.T. wrote the manuscript with contributions from all co-authors.

**Corresponding author**

Correspondence should be addressed to Vasily S. Minkov.

**Competing interests**

The authors declare no competing interests.

**Data availability**

The source data that support the findings of this study are provided with this paper.